\documentclass[aps,twocolumn,amsmath,amsfonts,nofootinbib,preprintnumbers,
  superscriptaddress,secnumarabic]{revtex4-1}
\usepackage[utf8]{inputenc}
\usepackage{color,graphicx,hyperref,amsmath,amssymb,multirow,slashed}
\usepackage[section]{placeins}
\hypersetup{colorlinks=true,linkcolor=blue,citecolor=magenta,filecolor=magenta,
  urlcolor=cyan}

\newcommand{\be}{\begin{equation}}
\newcommand{\ee}{\end{equation}}
\def\bsp#1\esp{\begin{split}#1\end{split}}
\def\bpm{\begin{pmatrix}}
\def\epm{\end{pmatrix}}

\def\lag{{\cal L}}
\def\sss{\scriptscriptstyle}
\def\as{\alpha_{\sss s}}

\def\dR{d_{\sss R}}
\def\dRbar{{\bar d}_{\sss R}}
\def\eR{\ell_{\sss R}}
\def\eRbar{{\bar e}_{\sss R}}
\def\Ll{L_{\sss L}}

\def\QL{Q_{\sss L}}
\def\QLbar{\bar Q_{\sss L}}
\def\uR{u_{\sss R}}
\def\uRbar{{\bar u}_{\sss R}}
\def\mLQ{m_{\sss LQ}}
\def\yoLL{{\bf y_{\sss 1}^{\sss LL}}}
\def\ydLL{{\bf y_{\sss 2}^{\sss RL}}}
\def\ytLL{{\bf y_{\sss 3}^{\sss LL}}}

\def\muR{\mu_{\sss R}}
\def\muF{\mu_{\sss F}}
\def\muzero{\mu_{\sss 0}}

\newcommand{\fa}{{\sc\small FeynArts}}
\newcommand{\fc}{{\sc\small FormCalc}}
\newcommand{\collier}{{\sc\small Collier}}
\newcommand{\fr}{{\sc \small FeynRules}}

\newcommand{\mg}{{\sc\small MG5\_aMC}}

\newcommand{\nloct}{{\sc\small NLOCT}}
\newcommand{\pwb}{{\sc\small Powheg-Box}}

\begin{document}
\title{Scalar leptoquark pair production at hadron colliders}

\author{Christoph~Borschensky}
\email{christoph.borschensky@uni-tuebingen.de}
\affiliation{Institute for Theoretical Physics, University of T\"ubingen,
  Auf der Morgenstelle 14, 72076 T\"ubingen, Germany}
\author{Benjamin~Fuks}
\email{fuks@lpthe.jussieu.fr}
\affiliation{Sorbonne Universit\'e, CNRS, Laboratoire de Physique Th\'eorique et
  Hautes \'Energies, LPTHE, F-75005 Paris, France}
\affiliation{Institut Universitaire de France, 103 boulevard Saint-Michel,
  75005 Paris, France}
\author{Anna~Kulesza}
\email{anna.kulesza@uni-muenster.de}
\affiliation{Institute for Theoretical Physics, WWU M\"unster, D-48149 M\"unster,
  Germany}
\author{Daniel~Schwartl\"ander}
\email{d\_schw20@uni-muenster.de}
\affiliation{Institute for Theoretical Physics, WWU M\"unster, D-48149 M\"unster,
  Germany}

\begin{abstract}
We revisit scalar leptoquark pair production at hadron colliders. Apart from QCD
contributions, we include the lepton $t$-channel exchange diagrams relevant
in the light of the recent $B$-flavor anomalies. We evaluate all contributions
at next-to-leading order in QCD and resum, in the threshold regime, soft-gluon
radiation at next-to-next-to-leading-logarithmic accuracy. All corrections are
found equally relevant. Our predictions consist of the most precise leptoquark
cross section calculations available to date and are necessary for the best
exploitation of leptoquark LHC searches.
\end{abstract}

\maketitle

{\it Introduction} --
Many extensions of the Standard Model (SM) predict the existence of scalar
leptoquarks~\cite{Pati:1974yy,Georgi:1974sy,Fritzsch:1974nn,Dimopoulos:1979es,
Senjanovic:1982ex,Schrempp:1984nj,Hewett:1988xc,Frampton:1989fu}, {\it i.e.}
scalar bosons coupling to a quark and a lepton simultaneously. Evidence for
their existence is consequently vastly searched for at the LHC. However, none
of the recent ATLAS~\cite{Aaboud:2019jcc,Aaboud:2019bye} and
CMS~\cite{Sirunyan:2018jdk,Sirunyan:2018ryt,Sirunyan:2018ruf,Sirunyan:2018vhk,Sirunyan:2018btu}
analyses find any hint for these leptoquarks, so that
their mass is now constrained to be larger than 1--1.5~TeV. Recently, scalar
leptoquarks have gained a significant interest as they may provide an
explanation~\cite{Bauer:2015knc,Hiller:2014yaa,Becirevic:2017jtw,Crivellin:2017zlb,Buttazzo:2017ixm,
Angelescu:2018tyl,Crivellin:2019dwb} for the $B$-meson anomalies~\cite{Lees:2012xj,Hirose:2016wfn,
Abdesselam:2019wac,Abdesselam:2019lab,Aaij:2015yra,Aaij:2017vbb,Aaij:2017uff,
Aaij:2019wad} and address~\cite{Dorsner:2019itg} the discrepancy between
theoretical predictions~\cite{Jegerlehner:2009ry} and experimental
measurements~\cite{Bennett:2006fi} of the anomalous magnetic moment of the muon
$(g-2)_\mu$. In this context, favored scenarios generally feature large
lepton-quark-leptoquark Yukawa couplings $y$.

The most stringent bounds originating from LHC direct searches for
leptoquark pair production and decay are extracted by assuming that leptoquarks
are solely produced via strong interactions. In other words, non-QCD diagrams
involving lepton $t$-channel exchanges of ${\cal O}(y^2)$ are neglected. In the
associated limit setting procedure, signal cross sections evaluated at
next-to-leading-order (NLO) accuracy in the strong coupling $\as$~\cite{Sirunyan:2018jdk,Sirunyan:2018ryt,Sirunyan:2018ruf,Sirunyan:2018vhk,Sirunyan:2018btu}, sometimes also supplemented by  
logarithmic threshold corrections~\cite{Aaboud:2019jcc,Aaboud:2019bye}, are used. Thus the predictions include contributions at ${\cal O}(\as^2)$ and ${\cal O}(\as^3)$, or possibly of higher order in $\as$, but are
independent of $y$~\cite{Kramer:2004df,Mandal:2015lca}. Bearing in mind the
$B$-anomalies and $(g-2)_\mu$ motivation, the limits may thus be incorrectly estimated.

In this paper, we perform for the first time a full NLO-QCD cross section
calculation for scalar leptoquark pair production at hadron colliders, in which
we include both the QCD and $t$-channel contributions. 
Hadronic production of heavy systems, which is the case considered here, inevitably probes partonic center-of-mass energies close to the production threshold given by twice the leptoquark mass $\mLQ$. In this limit, radiative corrections are dominated by soft-gluon emissions, manifesting themselves as large logarithmic terms that must be consistently resummed to all orders~\cite{Sterman:1986aj,Catani:1989ne,Contopanagos:1996nh,Catani:1996yz}. We report here threshold-resummed results at next-to-next-to-leading-logarithmic (NNLL) accuracy and showcase predictions obtained by matching them to our new NLO results.
 In the following, we first present the
considered theoretical framework and provide brief technical computational
details. We then show an illustrative selection of results that underlines how
all considered corrections affect the results in comparable and significant
ways. Our predictions, which are the most precise to date, are hence required to
derive limits consistently, in particular when assessing the influence of the
leptoquark Yukawa couplings.

\medskip

{\it Theoretical framework} --
We focus on a simplified model in which the SM is supplemented by several
species of scalar leptoquarks $S_1$, $\tilde S_1$, $R_2$, $\tilde R_2$ and
$S_3$. Inspired by standard naming conventions~\cite{Buchmuller:1986zs,
Dorsner:2016wpm}, these leptoquarks lie in the $({\bf 3},{\bf 1})_{\sss -1/3}$,
$({\bf 3},{\bf 1})_{\sss -4/3}$, $({\bf 3},{\bf 2})_{\sss 7/6}$,
$({\bf 3},{\bf 2})_{\sss 1/6}$ and $({\bf 3},{\bf 3})_{\sss-1/3}$
representations
of the SM gauge group respectively, and we target their Yukawa interactions
involving exactly one lepton and quark. The latter are collected in the
Lagrangian:
\be\bsp
 & \hspace*{-.2cm}\lag_{\rm int.} =
   {\bf y_{\sss 1}^{\sss RR}} \uRbar^c \eR^{\phantom{c}} S_1^\dagger
  +{\bf y_{\sss 1}^{\sss LL}} (\QLbar^c \!\cdot\! \Ll^{\phantom{c}}) S_1^\dagger
  +{\bf \tilde{y}_{\sss 1}^{\sss RR}} \dRbar^c \eR^{\phantom{c}} \tilde S_1^\dagger
 \\ &\ \hspace*{-.2cm} +
   {\bf y_{\sss 2}^{\sss LR}} \eRbar^{\phantom{c}} \QL^{\phantom{c}}R_2^\dag
  +{\bf y_{\sss 2}^{\sss RL}} \uRbar^{\phantom{c}}(\Ll^{\phantom{c}}\!\cdot\! R_2)
  +{\bf \tilde{y}_{\sss 2}^{\sss RL}} \dRbar^{\phantom{c}}(\Ll^{\phantom{c}}\!\cdot\! \tilde R_2)
 \\ &\ \hspace*{-.2cm}+
    {\bf y_{\sss 3}^{\sss LL}} \big(\QLbar^c\!\cdot\!\sigma_k \Ll^{\phantom{c}}\big)
      (S_3^k)^\dag
   + {\rm H.c.}\ .
\esp\label{eq:lag}\ee
In this expression, all flavor indices are suppressed for clarity, $\sigma_k$
stands for the Pauli matrices and the dot for the invariant
product of two fields lying in the (anti)fundamental representation of $SU(2)$. The $\QL$
and $\Ll$ spinors denote the SM weak doublets of left-handed quarks and
leptons, and $\uR$, $\dR$ and $\eR$ are the corresponding weak singlets.
Moreover, the ${\bf y/\tilde{y}}$ couplings are $3\times 3$ matrices in the
flavor space, the first index of any element $y_{ij}/\tilde y_{ij}$
referring to the quark generation and the second one to the lepton generation in
the gauge basis.

The calculations reported in this work concern scalar leptoquark pair production
and include fixed order contributions at leading order (LO) and NLO in QCD.
In contrast with previous work~\cite{Kramer:1997hh, Kramer:2004df,
Mandal:2015lca, Dorsner:2018ynv}, we not only consider the QCD components
at ${\cal O}(\as^2)$ and ${\cal O}(\as^3)$, but also include the $t$-channel
lepton exchange contributions at ${\cal O}(y^4)$ and ${\cal O}(y^4\as)$ as well as
the ${\cal O}(y^2 \as)$ and
${\cal O}(y^2 \as^2)$ interference of the $t$-channel diagrams with the QCD ones. The full NLO-accurate predictions are collectively
coined ``NLO w/ $t$-channel'' in the following, in contrast to the pure QCD ones
that we refer to as the ``NLO-QCD'' predictions. The NLO w/ $t$-channel cross sections are then additively 
matched with the resummed NNLL soft-gluon contributions, resulting in cross section predictions at NLO  w/ $t$-channel+NNLL accuracy. 
Threshold resummation is performed in Mellin space (see {\it e.g.}~\cite{Catani:1996yz})  and involves one-loop matching coefficients~\cite{Beenakker:2016gmf}.

To ensure the correctness of the results, we perform the calculations in two
independent ways: We first implement the above model into
\fr~\cite{Alloul:2013bka}, which we jointly use with
\nloct~\cite{Degrande:2014vpa} and \fa~\cite{Hahn:2000kx} to
renormalize the bare Lagrangian of eq.~\eqref{eq:lag} at ${\cal O}(\as)$. We
then generate a UFO model file~\cite{Degrande:2011ua} that we use to evaluate
fixed-order LO and NLO predictions within the \mg\
framework~\cite{Alwall:2014hca}. The latter are cross-checked with results
obtained within the \pwb\ framework~\cite{Alioli:2010xd}, in which
we input virtual corrections calculated with the \fa, \fc~\cite{Hahn:1998yk}
and \collier~\cite{Denner:2016kdg,Denner:2002ii,Denner:2005nn,Denner:2010tr} packages.
The NNLL corrections are evaluated with two independent in-house Monte Carlo codes.

\medskip

{\it Scalar leptoquark pair production at the LHC} --
We present selected predictions for scalar leptoquark pair production at the
13~TeV LHC for the three most commonly discussed types of scalar leptoquarks in
the context of the flavor anomalies: the $SU(2)_L$ singlet state $S_1$ (denoted
by $S_1^{\sss (-1/3)}$ due to its electric charge of $-1/3$), doublet state
$R_2$ and the triplet state $S_3$. More specifically, in the last two cases, we
consider the pair production of the $R_2$ mass eigenstate of electric charge of
$5/3$ (denoted by $R_2^{\sss (5/3)}$) and the one of the $S_3$ mass eigenstate
of electric charge of $-4/3$ (denoted by $S_3^{\sss (-4/3)}$). In all
our calculations, we treat the leptoquark mass $\mLQ$ as a free parameter and
assume the CKM matrix to be diagonal. While the determination of a scenario
compatible with flavor constraints and $Z$-pole observables is
desirable~\cite{Arnan:2019olv}, this goes beyond the scope of this study.
We consider instead benchmarks motivated by
ref.~\cite{Angelescu:2018tyl}.
The values of the Yukawa couplings found in this
study were obtained in a fit to low-energy observables and did not involve
constraints from direct searches for leptoquarks at the LHC. Given that the
description of lepton flavor university-violating observables involves both
leptoquark couplings and masses, optimally one should aim at a global fit based
on direct and indirect constraints, in which case the calculations presented in
this work will play a crucial role. For $S_1 S_1^*$ production, we
adopt a minimal
flavor ansatz for the leptoquark Yukawa couplings, $(\yoLL)_{\sss 22} = -0.15$
and $(\yoLL)_{\sss 32} =3$ with all other $\yoLL$ elements set to 0. For
$R_2R_2^*$ production, we similarly consider as the only non-vanishing coupling
$(\ydLL)_{\sss 22}=1.5$, a value still allowed by direct exclusion
bounds~\cite{Angelescu:2018tyl}, while for $S_3S_3^*$ production, we adopt
$(\ytLL)_{\sss 22}=-(\ytLL)_{\sss 32}$, keeping the actual coupling value free and
setting all other couplings to 0.

Our results are obtained by convoluting the partonic results with two different
sets of parton distribution functions (PDFs),
NNPDF3.1~\cite{AbdulKhalek:2019bux} and CT18~\cite{Hou:2019efy}. Unless stated
otherwise, NLO sets are employed for NLO-QCD and NLO w/ $t$-channel predictions,
while NNLO sets are used for NLO+NNLL calculations.
We set the renormalization ($\muR$) and factorization ($\muF$) scales
equal to a common value $\mu = \muR = \muF$. The central scale choice $\mu =
\muzero$ is fixed to $\muzero = \mLQ$, and scale uncertainties are estimated by
varying $\mu$ by a factor of 2 up and down. 

\begin{figure*}
  \centering
  \includegraphics[width=0.48\textwidth]{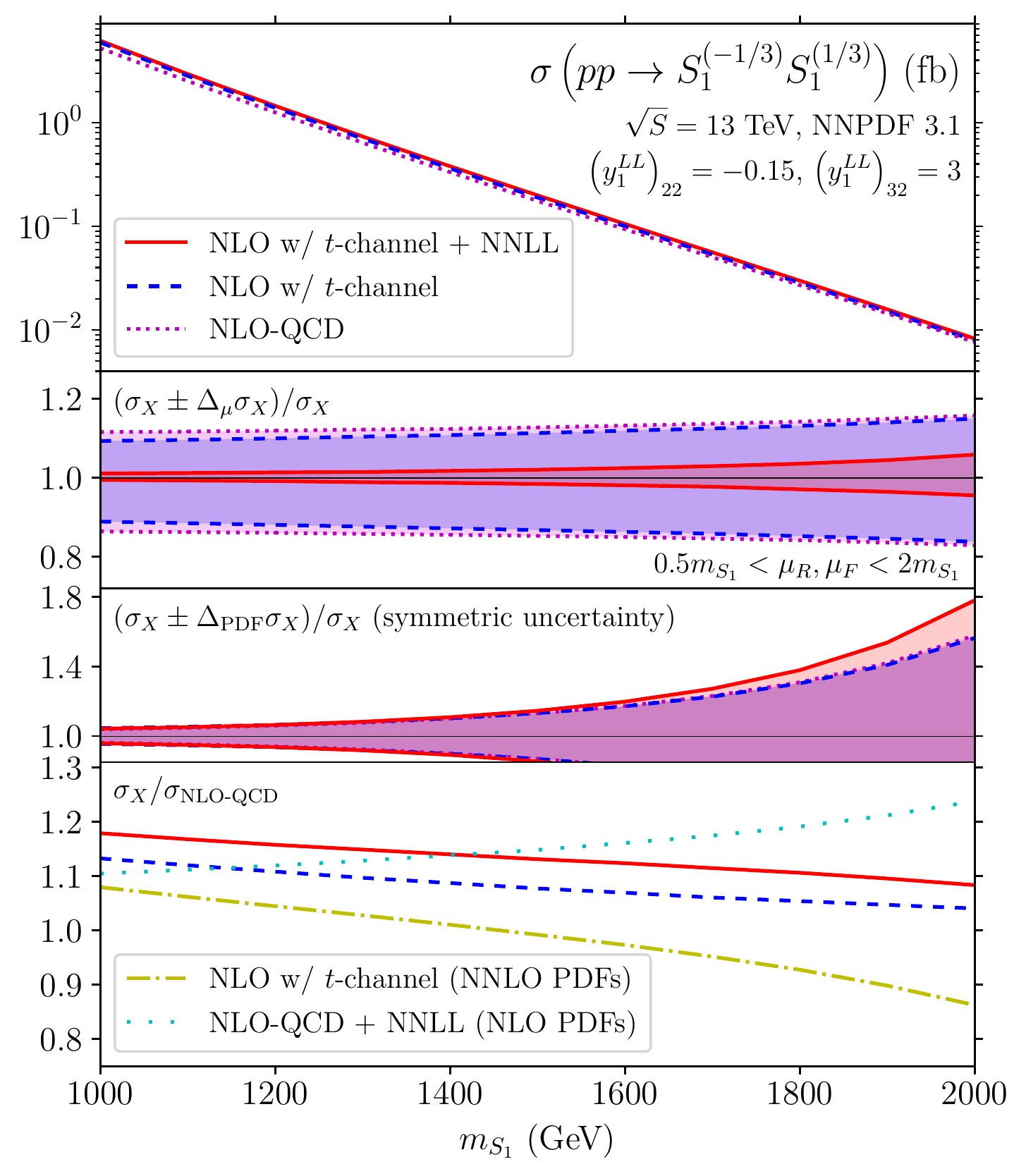}
  \includegraphics[width=0.48\textwidth]{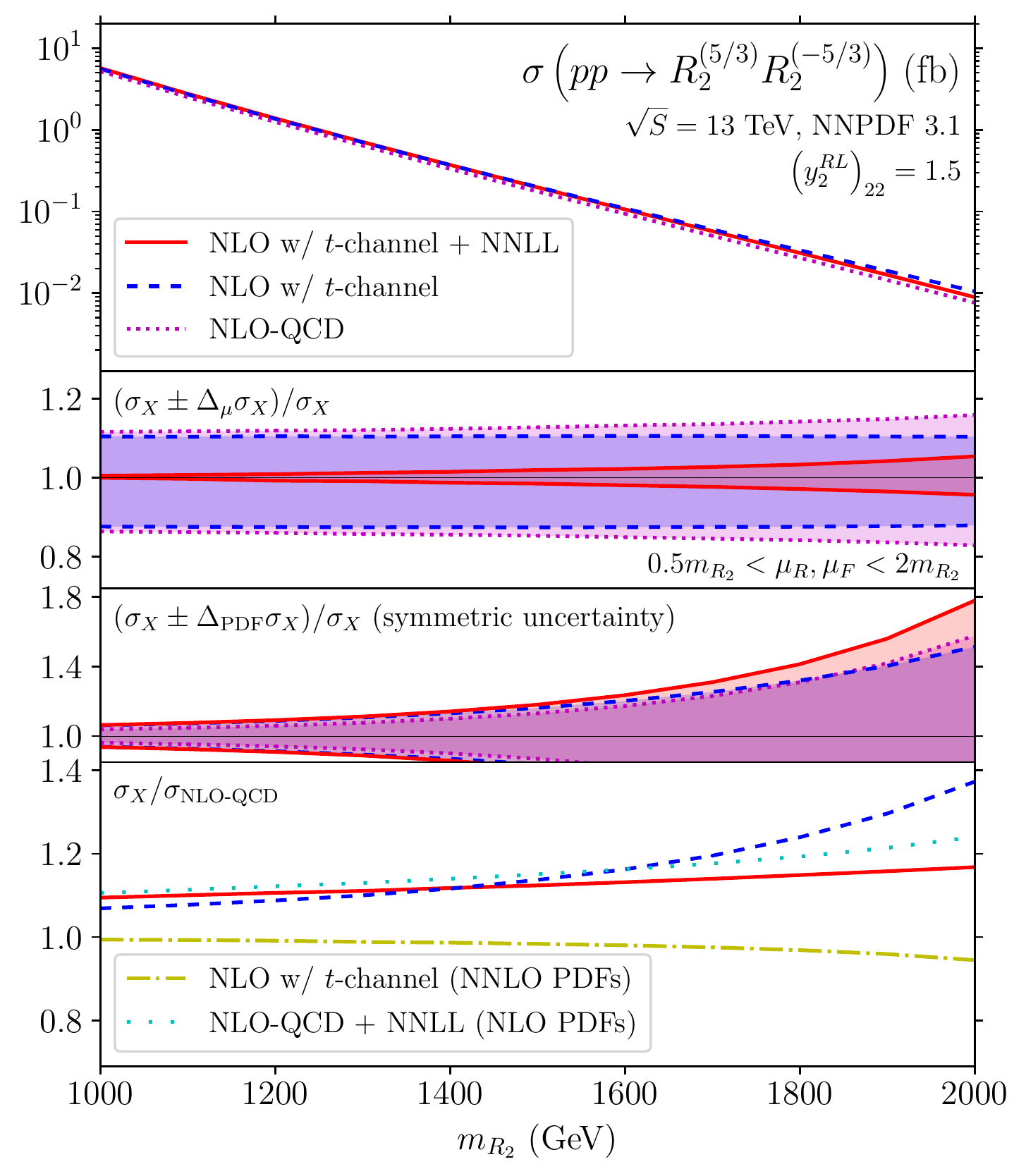}\\
  \includegraphics[width=0.48\textwidth]{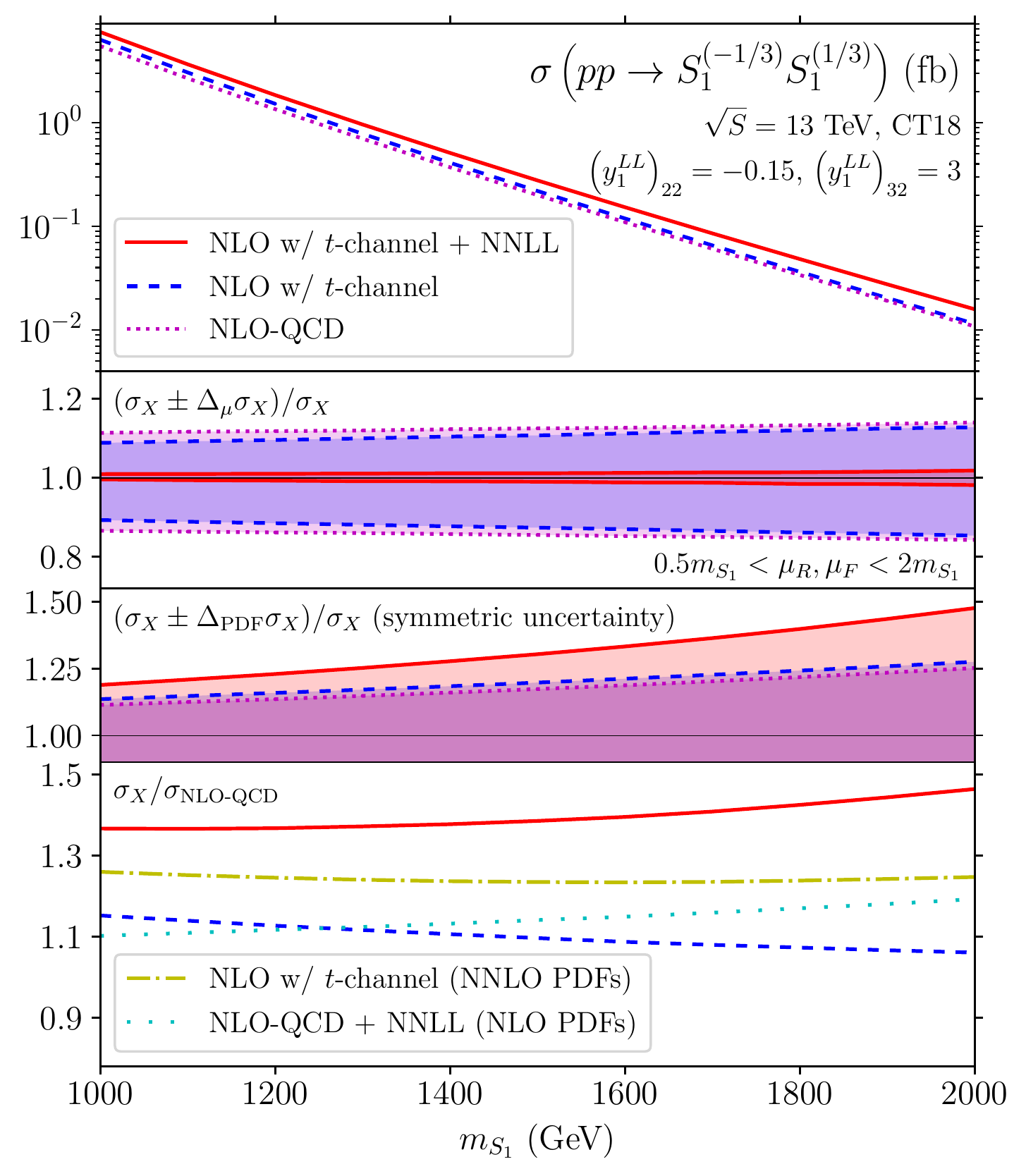}
  \includegraphics[width=0.48\textwidth]{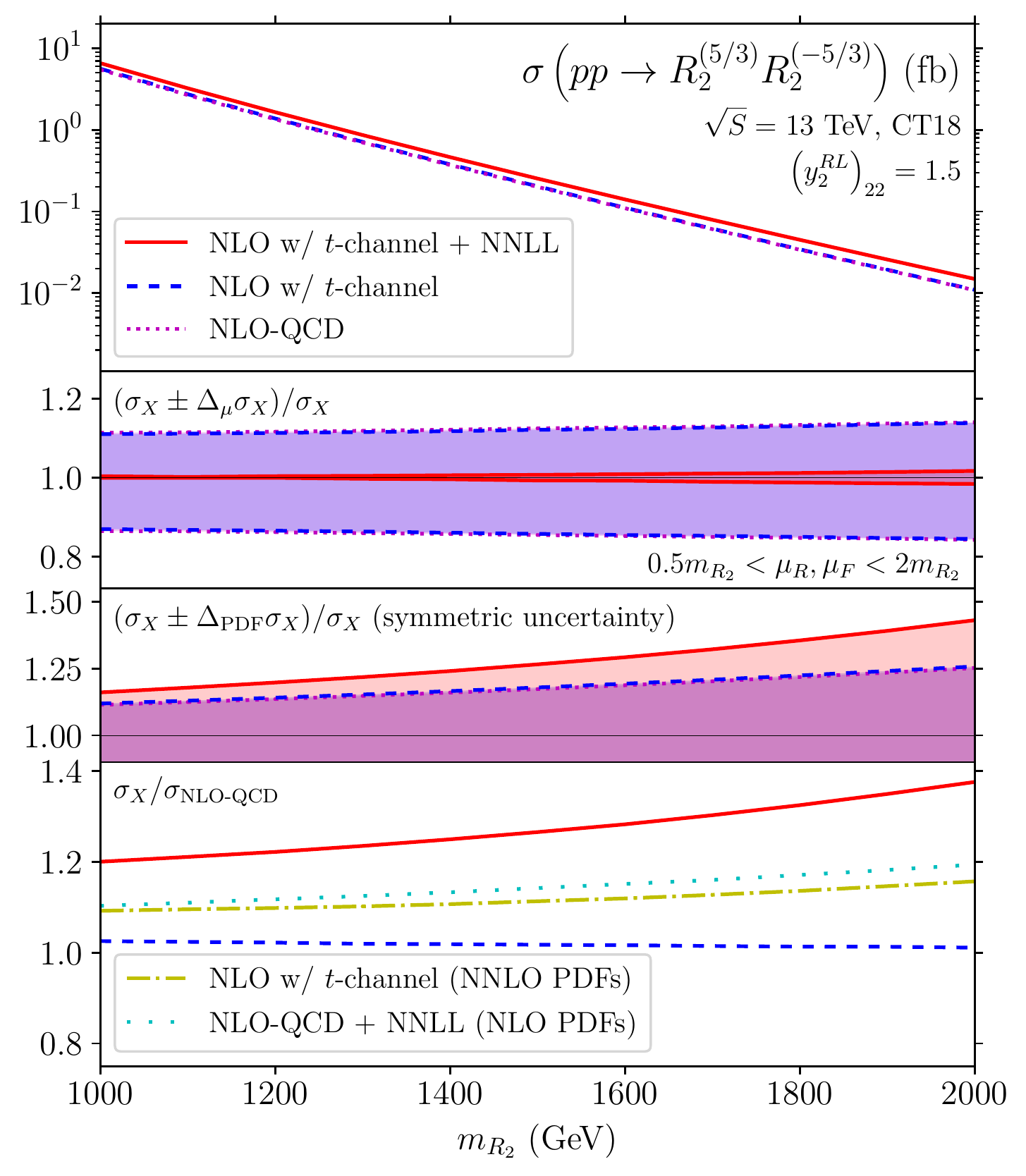}
  \caption{$S_1^{\sss(-1/3)} S_1^{\sss(1/3)}$ (left) and $R_2^{\sss(5/3)}
    R_2^{\sss(-5/3)}$ (right) production at the 13~TeV LHC, using the NNPDF3.1
    (upper row) and CT18 (lower row) PDF sets. In the top panels of the subfigures,
    we present cross section predictions at the NLO-QCD (magenta dotted), NLO w/
    $t$-channel (blue dashed) and NLO w/ $t$-channel+NNLL (red solid) accuracy.
    The associated scale and PDF uncertainties are also displayed (middle panels).
    In the lower panels, we show ratios of the NLO w/
    $t$-channel, NLO w/ $t$-channel+NNLL, NLO w/ $t$-channel calculated using
    NNLO PDFs (olive dash-dotted) and NLO-QCD+NNLL (turquoise dotted) results to
    the NLO-QCD cross section.}
 \label{fig:xsection}
\end{figure*}

\begin{figure*}[t]
  \centering
  \includegraphics[width=0.48\textwidth]{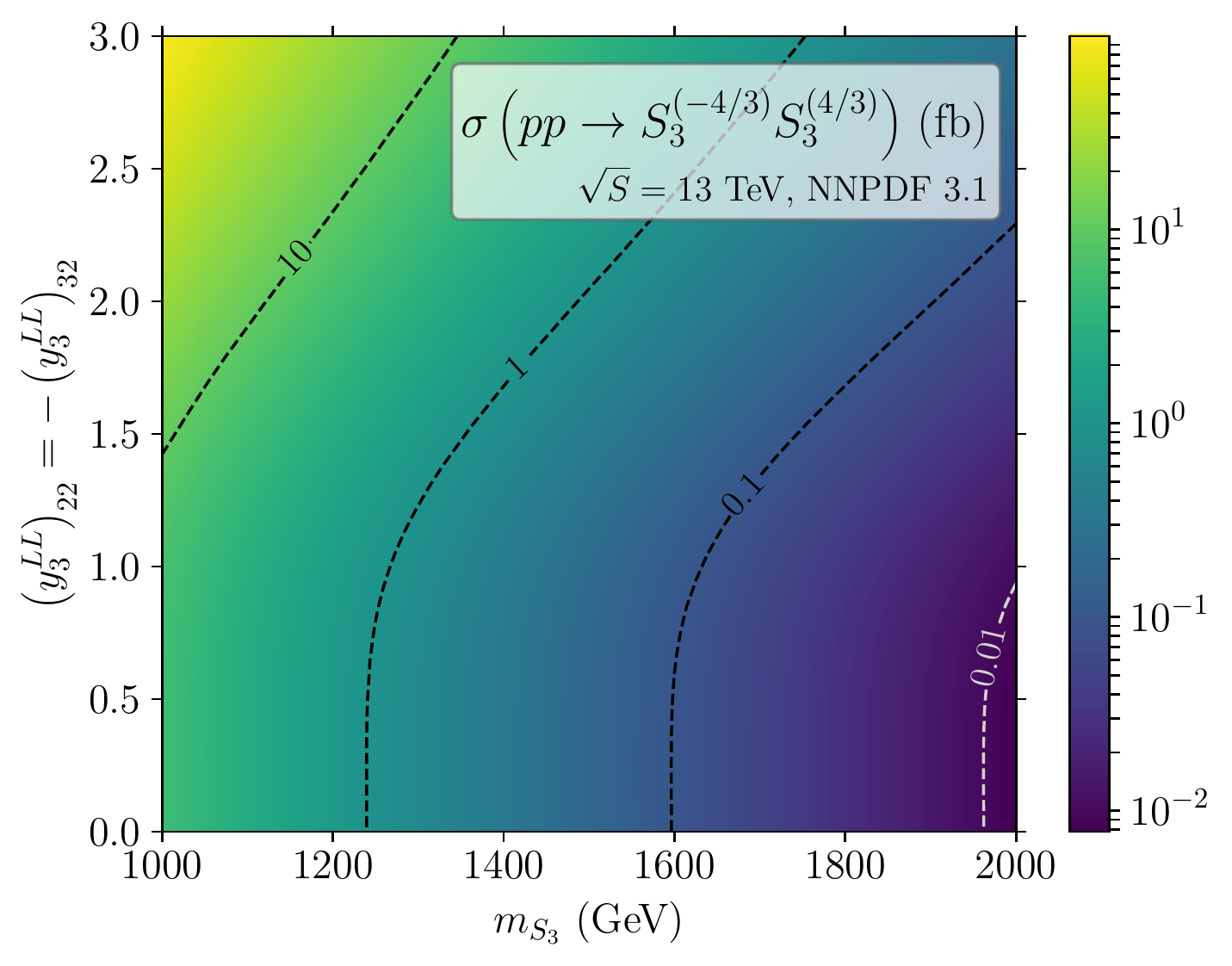}
  \includegraphics[width=0.48\textwidth]{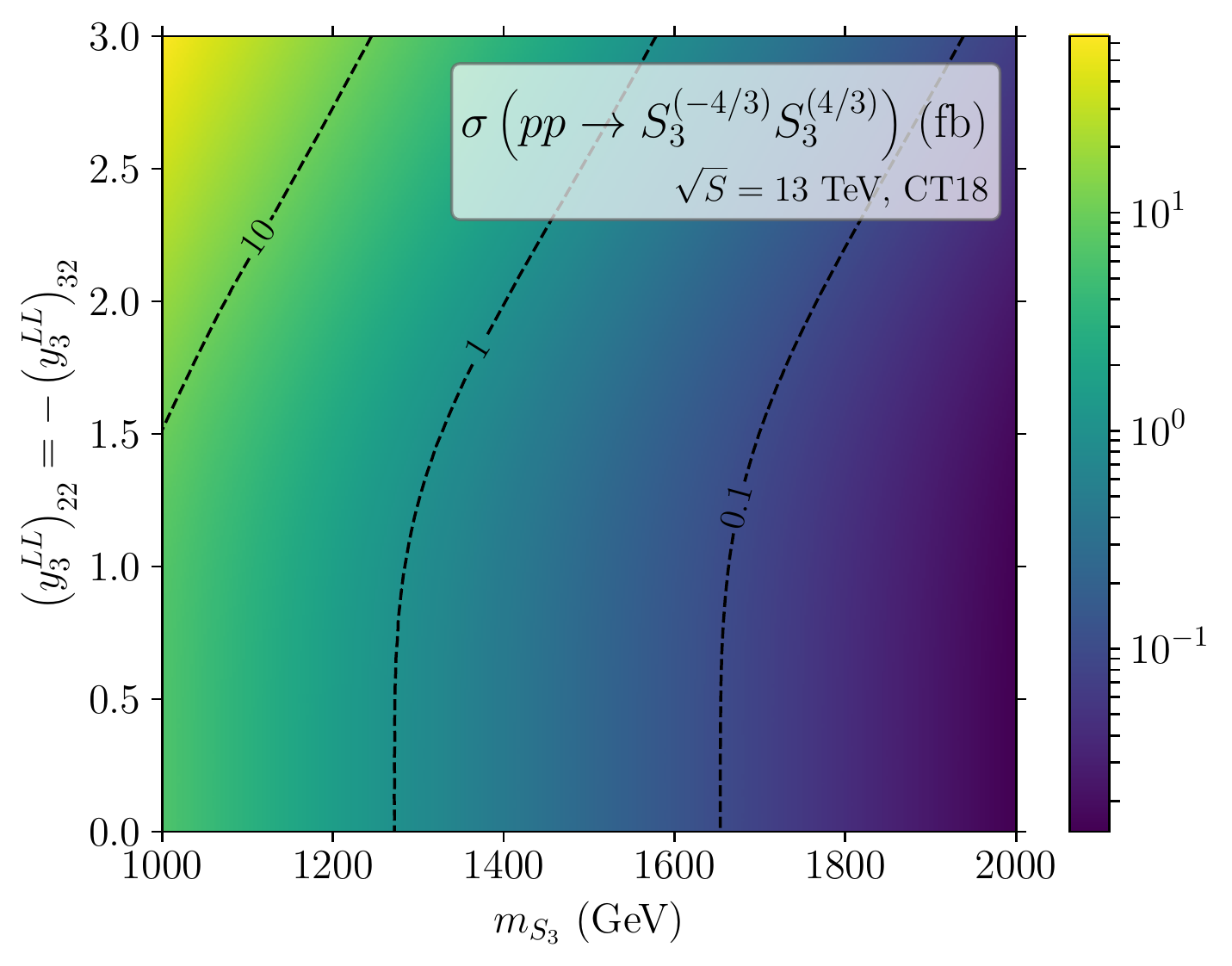}
  \caption{NLO w/ $t$-channel+NNLL total cross section for $S_3^{\sss(-4/3)}
    S_3^{\sss(4/3)}$ production at the 13~TeV LHC as a function of the $S_3$
    mass $\mLQ\!=\!m_{\sss S_3}$ and the Yukawa couplings $(\ytLL)_{\sss 22}
    \!=\! -(\ytLL)_{\sss 32}$ (all other Yukawa couplings being set to 0). We
    present predictions obtained with the NNPDF3.1 (left) and CT18 (right) PDF
    set.}
  \label{fig:xsectionS3}
\end{figure*}

In fig.~\ref{fig:xsection}, we present cross section predictions for
$S_1^{\sss (-1/3)} S_1^{\sss (1/3)}$ (left column) and $R_2^{\sss (5/3)}
R_2^{\sss (-5/3)}$ (right column) production, both for the NNPDF3.1 (upper row) and CT18 (lower row)
parton densities. We estimate the relative importance of the various corrections
studied in this work with respect to NLO-QCD predictions, and assess the
size of the scale and PDF uncertainties. Comparing the four subfigures, we observe
that depending on the process, the PDFs, the magnitude of the Yukawa
couplings, and $\mLQ$, the considered corrections can influence the predictions
in different, often contrasting, ways.

Although providing a positive correction, the size of the $t$-channel contributions depends
very differently on $\mLQ$ for the two processes (blue dashed curves). On the
contrary, NNLL effects, which we estimate through the ratio of the NLO-QCD+NNLL
to the NLO-QCD cross sections both calculated with the same NLO PDF set
(turquoise dotted curves), are independent of the process and PDF choice. As
expected, this ratio is bigger than 1 for all $\mLQ$ considered, and grows with
increasing $\mLQ$, {\it i.e.}, approaching the production threshold.

This behavior is vastly modified by the interplay of $t$-channel contributions, PDF effects and soft-gluon corrections, all entering the NNLL results matched with NLO w/ $t$-channel (red solid curves).
The effect of evaluating the NLO w/ $t$-channel cross sections with NNLO PDF sets instead of NLO sets is illustrated by the difference between the corresponding ratios to NLO QCD predictions (blue dashed {\it vs.}~olive dash-dotted curves).
For the NNPDF3.1 PDF set (upper row), this effect diminishes the cross sections, offsetting the increase stemming from the NNLL contributions. As a consequence, the NLO w/ $t$-channel+NNLL results deliver a
positive correction of about 10--20\% with respect to the NLO-QCD predictions
for $\mLQ\in[1, 2]$~TeV. Similarly for the NLO w/ $t$-channel result, the full NLO w/ $t$-channel+NNLL correction
exhibits the opposite behavior with increasing $\mLQ$ in the $S_1$ (upper left) and $R_2$ (upper right) cases.

When CT18 PDFs are used instead (lower row), the corrections are larger and reach a magnitude of about 20--50\%, the impact this time increasing with $\mLQ$ for both processes. Furthermore, in the NNPDF3.1 case, various contributions to the total correction are
often much bigger than the correction itself. For example, the correction due to
including $t$-channel diagrams reaches up to 40\% of the NLO-QCD result for
the pair production of 2~TeV $R_2$ leptoquarks, whereas the complete
NLO w/ $t$-channel+NNLL one is only of about 20\%. In contrast, results
obtained with CT18 densities exhibit the opposite behavior, the cross sections
being typically enhanced when switching from NLO to NNLO PDFs. The $t$-channel
and soft-gluon resummation pieces are of comparable size and thus equally
contribute to the combined correction. Therefore a precise knowledge
of the cross section requires calculating all classes of corrections.

In the middle panels of the four subfigures of fig.~\ref{fig:xsection}, we focus
on scale and PDF uncertainties. We
distinguish the impact of scale variations (second panel) from the one
originating from the PDF determination (third panel). Our results show that
soft-gluon resummation leads to a significant reduction of the scale
uncertainties from around 10\% (for the NLO predictions) to about 1--2\% for
$\mLQ$ values ranging up to slightly above the current exclusion limits. The
reduction might however be underestimated due to the chosen method for scale
uncertainty evaluation. This calls for a more comprehensive study.
Correspondingly, the total theoretical error for our final NLO w/
$t$-channel+NNLL predictions is dominated by its PDF component. The size of
the PDF error is however strongly dependent on the PDF choice. For instance,
results derived with NNPDF3.1 exhibit, for $\mLQ\!\sim\!1$~TeV, PDF errors
smaller or comparable in magnitude to the size of the perturbative corrections,
whilst at higher $\mLQ$ values, the PDF error becomes significantly bigger. In
comparison, PDF errors obtained with the CT18 set are larger for
small $\mLQ$ values, but do not grow as quickly for higher masses. Still, the
PDF errors turn out to be of the same order as the full perturbative corrections
for large $\mLQ$ values. Those large PDF errors at high masses hence obscure
the accuracy of the predictions. However, as more LHC data will be
analyzed, one can expect a substantial improvement of the PDF knowledge, in
particular in the large Bjorken-$x$ regime, so that the PDF errors associated
with predictions relevant for high-mass system production will be significantly
reduced.

In fig.~\ref{fig:xsectionS3}, we calculate the NLO w/ $t$-channel+NNLL total
cross section for $S_3^{\sss (-4/3)}S_3^{\sss(4/3)}$ production, and study its
dependence on the leptoquark mass and Yukawa coupling strength. We consider both
the NNPDF3.1 (left) and CT18 (right) PDF sets. At small values
of the Yukawa coupling, the dominant production mechanism is QCD driven so that
the cross section solely depends on $\mLQ$. On the contrary, as the coupling
approaches 1, the $t$-channel contributions become more relevant and the
total rate significantly increases. This behavior is mostly independent of the
chosen PDF set, CT18 predictions being slightly less sensitive to the
$t$-channel diagrams.

\medskip

{\it Summary} --
We have significantly advanced the precision of scalar leptoquark
pair production cross section computations. First, we have included all
contributions to the process, both the QCD ones and those involving
the $t$-channel exchange of a lepton, at NLO QCD. Second, we have resummed
soft-gluon radiation in the threshold regime to NNLL accuracy.

The $t$-channel contributions, threshold resummation, the adopted
parton densities and benchmark scenario (in particular when the leptoquark
Yukawa couplings are taken as large as suggested by the recent $B$-anomalies)
importantly affect the total rates, in potentially contrasting and sizable ways.
This emphasizes the necessity of including all contributions whose calculation has
been pioneered in this work. While the perturbative series exhibits smaller
scale uncertainties, the precision of the predictions is limited by the poor
PDF knowledge in the large Bjorken-$x$ regime relevant for the
production of high-mass systems. In light of our findings, we recommend the
usage of NLO w/ $t$-channel+NNLL cross sections, to be taken together with the correspondingly
reduced scale uncertainties and PDF errors extracted from the envelope
spanned by computations, left for future work, performed with different PDF
sets. This follows the strategy outlined in various recommendations for LHC cross section calculations~\cite{Botje:2011sn, Dittmaier:2011ti,  Kramer:2012bx, Fuks:2013vua, Borschensky:2014cia}.
The computer codes used in this work are available upon request.



{\it Acknowledgments} --
We are grateful to V.~Hirschi, O.~Mattelear and H.S.~Shao for their help with
\mg\ technical issues related to mixed-order computations, as well as to
M.~Kr\"amer for insightful discussions, and D.~Be\v{c}irevi\'c, D.~Guadagnoli
and R.~Ruiz for useful comments on the
manuscript.  This work has been supported in part by the DFG grant
KU3103/2. We furthermore acknowledge support by the state of Baden-W\"urttemberg through bwHPC
and the DFG grant INST 39/963-1 FUGG (bwForCluster NEMO).

\bibliographystyle{utphys}
\bibliography{biblio}
\end{document}